\documentclass[aps,twocolumn,showpacs]{revtex4}
\usepackage{amsmath}

\begin{document}

\title{Quantum Process Tomography on vibrational states of atoms in an
Optical Lattice}
\author{S. H. Myrskog, J. K. Fox, M. W. Mitchell and A. M. Steinberg}
\affiliation{Dept. of Physics, University of Toronto, 60 St. George St. Toronto,Ont.,
Canada, M5S 1A7}

\begin{abstract}
Quantum process tomography is used to fully characterize the evolution of
the quantum vibrational state of atoms. Rubidium atoms are trapped in a
shallow optical lattice supporting only two vibrational states, which we
charcterize by reconstructing the 2x2 density matrix. Repeating this process
for a complete set of inputs allows us to completely characterize both the
system's intrinsic decoherence and resonant coupling.
\end{abstract}

\maketitle

In recent years, there have been remarkable advances in directly controlling
and observing the dynamics of individual quantum systems in a variety of
domains. \ This degree of control of microscopic systems is one of the
technological advances underlying the myriad proposals for realistic
quantum-information processing systems. \ For instance, a number of quantum
computing proposals rely on atoms trapped in an optical lattice \cite%
{lattice_comp99Brennen,Jaksch99-QC}, a system in which a great deal of work
has investigated coherent centre-of-mass motion\cite%
{Wavepacket-Ertmer00,tunnel_resonance_Raithel02,Raithel02feedback}, full
characterisation of spin states\cite{Klose-recon_dm01}, loading of
individual atoms into lattice sites \cite%
{patternedlattice,Bloch-collapse02,Weiss03-initializelattice}, and coherent
interactions between atoms\cite%
{lattcomp_Zoller00,Cphase_Deutsch02,Collisions03}. \ Here we demonstrate a
technique for completely reconstructing the quantum state of motion of an
atom trapped in a lattice well. \ By performing this density-matrix
reconstruction for a complete set of input states, we are able to completely
characterize the quantum evolution of the system (the ``superoperator''),
including decoherence.

The development of a quantum computer will rely on the success of
error-correction to reduce errors to an acceptable level\cite%
{QCThreshold1,QCThreshold2,QCThreshold3,QCThreshold4}. \ A quantum computer
is extremely vulnerable to errors and decoherence. \ Experimental
characterization of both the decoherence and the operations will be required
in order to implement quantum error correction \cite%
{Shor-QEC95,Steane96-QEC,Zanardi97,LidarDFS98}. \ An arbitrary operation may
be characterized using quantum process tomography\cite%
{Chuang-Nielson01,Poyatos_quantum-process97}, or QPT. \ The result of QPT is
the superoperator, a positive, linear map from density matrices to density
matrices, which governs the evolution of the density matrix for the
operation. \ Unlike a propagator, the superoperator allows for non-unitary
evolution of the system, thoroughly characterizing decoherence, relaxation
and loss in a system. \ From the superoperator one can determine the types
of errors which occur and develop procedures to reduce or eliminate errors,
without requiring a priori assumptions about the underlying physical
mechanism causing the errors\cite{Knill97}.

QPT\ has recently been demonstrated using spins in a NMR system\cite%
{Childs-Leung}, the polarization of single photons\cite{ancilla-assisted_QPT}
and a singlet state filter for photon pairs\cite{Mitchell03}. \ \ QPT is
performed by preparing a complete set of input density matrices, subjecting
each to the operation being tested, and measuring the resultant output
density matrices. \ Due to the linearity of quantum mechanics, QPT of a
process on an N dimensional system can be achieved by sending in $N^{2}$
linearly independent density matrices (alternatively, it has recently been
shown that one can use a single state in a larger Hilbert space as the input%
\cite{ancilla-assisted_QPT,Tomo-operations01}.) \ 

In this experiment we perform quantum process tomography using the motional
states of atoms trapped in the potential wells of a 1-D optical lattice. \ \
We examine processes which are independent of the electronic state of the
atom and are only dependent upon the motional states of the atom. \ The
measurements are insensitive to the long-range degrees of freedom and
effectively trace over the quasimomentum in the Bloch state picture (or
equivalently the well index in the Wannier state picture). \ We use a
shallow 1-D lattice which only supports 2 bound bands, which we label as
ground ($\left| 0\right\rangle $) and excited ($\left| 1\right\rangle $). \
The lattice is vertically oriented, causing all atoms in higher energy,
classically unbound states to quickly fall out of the lattice and become
spatially separated from atoms which remain in bound states of the lattice
(the Landau-Zener tunneling rates from the 3 lowest energy bands are 3$\cdot
10^{-7}$, 14.5 and 1150 per second in increasing order). \ A typical
sequence, from state preparation in the lattice to measurement, lasts 20 ms.

We begin by cooling and trapping Rubidium-85 atoms in a standard vapour cell
MOT to a temperature of 7 $\mu K$ with an rms radius of approximately 1 mm.
\ During the optical molasses stage we turn on an optical lattice in a
vertical orientation. \ The optical lattice is created by interfering two
laser beams which are detuned 30 GHz below the Rb D2 resonance at 780.03 nm.
\ Each beam travels through an acousto-optic modulator, each of which is
driven by a function generator, providing control of the relative phase
between the lattice beams.\ By modulating the phase of one of the lasers the
lattice may be displaced by up to a lattice spacing within one microsecond.
\ The beams are superposed after the acousto-optic modulators, and
co-propagate to the vacuum chamber with orthogonal polarizations, to reduce
phase fluctuations between the beams which would impress noise on the
lattice. \ The beams are separated on a polarizing beam splitter near the
MOT. \ The polarization of one beam is then rotated such that the beams have
parallel polarizations in the MOT. \ \ The beams have an angle of 50 degrees
between them, creating an optical lattice with a lattice constant of $L=0.93$
microns. \ The depth of the lattice is controlled by the intensity and
detuning of the beams, and is chosen to be $18$ $E_{r}$ (where $%
E_{r}=h^{2}/8L^{2}m=h\cdot 690$ Hz is the effective recoil energy of the
lattice) at which depth it contains two bound states. \ The energy
separation between the states is $\hbar \cdot 2\pi \cdot 5.0$ kHz. \ The
scattering rate from the lattice beams is on the order of 4 Hz, which is
insignificant on the timescale of the experiment. \ 

The population in each band of the lattice is determined by adiabatically
decreasing the lattice potential\cite{Jessen-adiab95,Mlynek97}. \ In order
to satisfy the adiabatic criterion we must decrease the potential slower
than $h$ $\cdot 10^{8}/s^{2}$, whereas the fastest decrease we use is $%
h\cdot 4.1\cdot 10^{6}/s^{2}$ with non-adiabatic effects appearing if the
turn-off is faster than $h\cdot 1.4\cdot 10^{7}/s^{2}$. \ As the depth of
the wells decreases the energy bands gradually get closer to the top of the
potential. \ Once an energy band becomes classically unbound, then all the
atoms in that band accelerate downwards due to gravity. \ Since each state
becomes unbound at a different time, each band becomes mapped into a
different location in space. \ The spatial distribution can be recorded on a
CCD camera by fluorescence imaging. \ Figure 1a shows a sample spatial
distribution after ramping down the potential over a time of 45
milliseconds. \ Alternatively, the lattice depth may be quickly, but still
adiabatically, lowered to a depth of 9 $E_{r}$. \ At this depth the ground
state has a Landau-Zener-limited lifetime of 250 milliseconds while the
excited state has a lifetime of 0.5 milliseconds. \ The excited state atoms
quickly escape the lattice while ground state atoms remain trapped. \ After
holding the depth constant for some time (typically about 20 ms), the
relative populations can be determined by fluorescence imaging, as shown in
Figure 1b. \ The beams creating the lattice have a Gaussian shape, with rms
width of 3 mm, causing the lattice to be shallower as we move farther away
from the center. \ The spatial variation of the lattice depth causes atoms
at the edge of the lattice to tunnel out sooner than atoms near the center,
resulting in the curved clouds seen in Figure 1a. \ \ To reduce broadening
effects due to inhomogeneous well depths we integrate the flourescence
signal only over the central 600 $\mu $m of the cloud where the potential
varies slowly.

A sample of ground state atoms is prepared by filtering out the excited
state atoms from the lattice. \ This is accomplished by reducing the well
depth to a depth of approximately 9 recoil energies, at which point only the
ground state is bound. \ The potential depth is held at this value for 3 ms,
long enough for almost all the excited state atoms to escape. \ The well
depth is then adiabatically increased back to the original depth, preparing
a sample of atoms with up to 95\% occupation of the ground state.\ 

To prepare a variety of initial states, we make use of our ability to
displace the lattice, and of the atoms' free evolution. \ \ Displacement of
the lattice is equivalent to a spatial translation of the atom cloud in the
lattice's reference frame, constituting a coherent coupling between the
energy eigenstates. In addition to coupling the two bound states, this
induces some transitions to unbound states, which are lost from the lattice.
Spatial translations change the coefficients of the states as described by
the following equations.

\begin{equation}
\begin{tabular}{|l|l|}
\hline
$\Delta x$ & 
\begin{tabular}{l}
$\left| 0\right\rangle \Longrightarrow c_{00}\left| 0\right\rangle
+c_{10}\left| 1\right\rangle +loss$ \\ 
$\left| 1\right\rangle \Longrightarrow -c_{10}^{\ast }\left| 0\right\rangle
+c_{11}\left| 1\right\rangle +loss$%
\end{tabular}
\\ \hline
\end{tabular}%
\end{equation}%
The coefficients $c_{00},c_{10}$ and $c_{11}$ are determined by displacing
the lattice with different initial state populations. \ A typical
displacement used during the measurement is 116 nm. \ For this displacement
we measure $c_{00}$,$c_{10}$ and $c_{11}$ to be 0.86(2), 0.50(2) and\
0.53(10) respectively, close to the theoretical values of 0.87,0.45 and
0.63. \ During a time period $\Delta t$ of free evolution in the lattice,
the ground and excited states acquire a relative phase shift of $\omega
\Delta t$, where $\omega $ is the oscillation frequency in the lattice. \
Using a combination of displacement and time delay we can prepare
superposition states with arbitrary relative phase.

State tomography\cite{Bigelow-staterecon} is performed by projecting the
unknown state onto a set of known states. \ We use a set of non-orthogonal
states $\left\{ \Phi _{1},...,\Phi _{4}\right\} =$ $\{\left| 0\right\rangle
,\left| 1\right\rangle ,\left| \theta _{x}\right\rangle ,\left| \theta
_{y}\right\rangle \}$ where\cite{Thew-Qudit02} $\left| \theta
_{x}\right\rangle =\cos \theta \left| 0\right\rangle +\sin \theta \left|
1\right\rangle $ and $\left| \theta _{y}\right\rangle =\cos \theta \left|
0\right\rangle -i\sin \theta \left| 1\right\rangle .$\ We project onto
states of the form $\cos \theta \left| 0\right\rangle +\sin \theta \left|
1\right\rangle $ by spatially displacing the trapping potential before
separating the resulting energy eigenstates. \ We choose our displacements
to be\ small in order to have negligible coupling to the higher energy,
unbound states. \ We use a displacement of L/8=116 nm, generating states
with an experimental value of $\theta \approx 0.5$ radians. \ The state $%
\left| \theta _{x}\right\rangle $ can be changed into state $\left| \theta
_{y}\right\rangle $ by a time delay of a quarter period after displacement.
\ State tomography is performed by measuring the projections $%
m_{i}=\left\langle \Phi _{i}\right| \rho \left| \Phi _{i}\right\rangle $ for
all states $\left| \Phi _{i}\right\rangle $. \ The resulting measurements $%
\left\{ m_{1},...,m_{4}\right\} $ are used to reconstruct the density matrix.%
\begin{equation}
\rho =\left( 
\begin{array}{cc}
m_{1} & \frac{(m3+i\ast m4)-m_{1}\cos ^{2}\theta -m_{2}\sin ^{2}\theta }{%
2\sin \theta \cos \theta } \\ 
\frac{(m3-i\ast m4)-m_{1}\cos ^{2}\theta -m_{2}\sin ^{2}\theta }{2\sin
\theta \cos \theta } & m_{2}%
\end{array}%
\right) 
\end{equation}

Process tomography is first performed on the free evolution of a quantum
state in the lattice for one period. \ This allows us to completely
characterize decoherence intrinsic to the lattice. \ To perform process
tomography we prepare a complete set of density matrices as input states. \
The four linearly independent density matrices we used correspond to a
ground state, prepared by filtering out the excited state population; a
coherent state with real coherence\ prepared by displacing a ground state; a
coherent state with imaginary coherence prepared by adding a quarter-period
time delay after displacement; and a mixed state. \ The mixed state can be
prepared by either skipping the filtering step, or by preparing a coherent
state and waiting 3 ms for it to decohere (see discussion below). \ 

QPT\ proceeds as follows: one of the input density matrices is prepared, and
characterized with state tomography as outlined above; the same state is
again prepared and allowed to freely evolve in the lattice for 200$\mu s$
(one oscillation period), and then state tomography is performed on the
resulting state. \ Figure 2 shows the projection of each input density
matrix onto the projection states, before and after the `operation'.

The super-operator $\mathcal{E}$ resulting from QPT can be expressed in a
number of ways. \ One common form is the operator sum representation,%
\begin{equation}
\rho _{out}=\mathcal{E}\left( \rho _{in}\right) =\sum\limits_{i}\widehat{A}%
_{i}\rho _{in}\widehat{A}_{i}^{\dagger }
\end{equation}%
where $\widehat{A}_{i}$\ are operational elements, often called Kraus
operators\cite{Kraus_Ops}, subject to the constraint $\underset{i}{\sum }%
\widehat{A}_{i}^{\dagger }\widehat{A}_{i}=I$\ . \ The Choi matrix\cite%
{Choi75,Choi_Proof-Leung} provides a straightforward procedure to obtain
experimental Kraus operators. \ The Choi matrix is defined as%
\begin{equation}
\underline{C}=\underset{i,j}{\sum }\left| i\right\rangle \left\langle
j\right| \otimes \mathcal{E}(\left| i\right\rangle \left\langle j\right| )
\end{equation}%
where $\left| i\right\rangle \left\langle j\right| $\ is an outer product of
basis states and $\mathcal{E}(\left| i\right\rangle \left\langle j\right| )$%
\ is the super-operator acting on the matrix given by the outer product $%
\left| i\right\rangle \left\langle j\right| $. Then%
\begin{equation*}
\mathcal{E}\left( \rho \right) =\underset{i,j}{\sum }C_{i,j}\rho _{i,j}
\end{equation*}%
where $C_{i,j}=\mathcal{E}(\left| i\right\rangle \left\langle j\right| )$ is
the i,jth 2$\times $2 submatrix of \underline{$C$} and $\rho _{i,j}$ is the
i,jth element of the density matrix. \ The eigenvalues and eigenvectors of 
\underline{$C$}\ can then be used to determine the canonical Kraus
operators, given by $\widehat{A}_{i}=\sqrt{\kappa _{i}}\widehat{k_{i}}$\
where $\kappa _{i}$\ is the ith eigenvalue and $\widehat{k_{i}}$\ is the
corresponding eigenvector written in matrix form. \ The matrix $\left|
i\right\rangle \left\langle j\right| $ is not necessarily a density matrix,
but can be written as a linear combination of the measured input density
matrices. \ Using a maximum-likelihood technique, we find the Choi matrix
which best predicts the measured output states given the measured input
states. \ This search is limited to physical, i.e., completely positive,
Choi matrices. \ The resulting Choi matrix is shown in Figure 3. \ From the
Choi matrix we find that the populations are preserved to within
experimental uncertainty while the coherences decay by 36 percent.\ 

The same data may be visually displayed using a Bloch sphere representation%
\cite{QuantuComp_Nielson,ancilla-assisted_QPT}, which has the advantage of
showing how any state on the surface of the Bloch sphere evolves into a new
state. \ Figure 4a shows the initial, undisturbed Bloch sphere before
evolution in the lattice, and Figure 4b shows the Bloch sphere after one
oscillation. \ The sphere becomes prolate, by contracting toward the z-axis
by 36\%, as expected. \ The Kraus operators are determined from the Choi
matrix. \ The most significant Kraus operators are found to be $\widehat{A}%
_{1}=0.90\widehat{I}+\widehat{R}_{1}$ and $\widehat{A}_{2}=-0.41\widehat{%
\sigma }_{z}+\widehat{R}_{2}$ where $\widehat{I}$ is the identity matrix, $%
\widehat{\sigma }_{z}$ is the z Pauli matrix and $\widehat{R}_{1},\widehat{R}%
_{2}$ are small remainders with magnitudes bound by $Tr[\widehat{R}%
_{i}^{\dagger }\widehat{R}_{i}]$ $\leq 0.03$. \ The other two Kraus
operators are insignificant on the scale of our experimental resolution,
also satisfying a similar bound. \ Kraus operators of the form $\widehat{I}$
and $\widehat{\sigma }_{z}$ are consistent with pure dephasing, as expected
for from either inter-well tunneling or inhomogeneous broadening.

The coherence time is thus found to be 555 $\mu s$ (2.78 periods); this is
shorter than the coherence time of 2 ms expected based on the width of the
excited band (the inter-well tunneling rate) and on the variation in well
depth across the finite Gaussian lattice beams. \ The number of oscillations
is however consistent with oscillations observed in other work\cite%
{tunnel_resonance_Raithel02,Haycock00,Raithel98}. \ It is believed that the
observed decoherence is caused by small-scale inhomogeneity, such as
fringes, in the lattice beams. \ The lattice beams have been spatially
filtered before travelling to the MOT. \ Unlike earlier experiments, we use
a two-level system, and anharmonicity is therefore not a factor in the
observed decoherence.

An example of an operation necessary for quantum information processing is a
single-qubit rotation. \ To demonstrate the applicability of process
tomography to characterizing such operations we attempt to perform a
rotation of the Bloch sphere by a method equivalent to a Rabi oscillation. \
We sinusoidally drive the displacement of the lattice at the trap frequency,
thereby coupling neighbouring states coherently. Process tomography is
performed after a single period of this drive. \ The lattice displacement is
driven by applying a sinusoidal phase modulation with an amplitude of $\pi
/9 $ radians to one of the lattice beams. \ The displacement is kept small
to ensure that coupling is predominantly between neighbouring states. \ \ We
test both a sine drive, $\Delta x(t)=x_{m}\sin (\omega _{0}t)$, and a cosine
drive, $\Delta x(t)=x_{m}(\cos (\omega _{0}t)-1)$, where $x_{m}=$ $26$ $nm$
and $\omega _{0}$ is the oscillation frequency in the lattice and the pulse
lasts from $t=0$ to $t=2\pi /\omega _{0}$. \ We again find the Choi matrix
from a maximum-likelihood model, but find the Bloch sphere representation to
be the most intuitive. Figures 4c and 4d show the resulting Bloch spheres,
which have rotations of 35.5 degrees and 36.4 degrees about the y-axis and
x-axis respectively. \ The Bloch spheres are rotated 90 degrees out of phase
with one another as expected for driving fields 90 degrees out of phase. \
The resulting shape is close to a sphere, but the radius has decreased in
all dimensions. \ In particular, the length of the semi-minor axis' for the
sinusoidal drive is 0.69 (while it should be noted that in the absence of
the coherent drive, it decayed to a value of 0.64). \ A simulation using a
truncated harmonic-oscillator model predicts a rotation of 35.0 degrees
about the y-axis.

We have introduced a new technique for determining the motional quantum
state of atoms in an optical lattice, and applied it to a demonstration of
quantum process tomography. \ We have extracted the ``superoperators'' fully
characterizing the action of several different operations on an arbitrary
state of atoms in the lattice, specifically, free evolution for one period
and two different resonant-coupling protocols.\ \ In this way, we have
characterized the intrinsic dephasing of the system over time, and the
effectiveness of single-qubit rotations induced by resonant modulation of
the lattice phase. We plan to extend these techniques to test the Markovian
approximation; to characterize and optimize bang-bang methods\cite%
{Lloyd-bangbang98,LidarBB} for removing inhomogeneous-broadening effects;
and to study the well-depth-dependence of the decoherence, investigating the
role of inter-well tunneling, Wannier-Stark transitions, and Bloch
oscillations. \ The procedure can be extended to higher-dimensional Hilbert
spaces, although the number of measurements required grows exponentially
with the dimensionality. \ Process tomography should prove essential for
tailoring error-correction protocols to the observed behaviour of particular
physical realisations of quantum-information systems\cite%
{Cory-incoher03,Cory-incoher03b}; it is reasonable to expect that such
system-by-system tailoring may prove necessary if error thresholds on the
order of $10^{-4}$ or $10^{-5}$ are ever to be reached\cite%
{QCThreshold1,QCThreshold2,QCThreshold3,QCThreshold4}. More generally, it is
the only method to permit complete characterisation of the evolution of open
quantum systems, and should play a central role in the toolbox for control
and study of individual quantum systems, whether for quantum-information
processing or for other applications.

We\ would like to thank Daniel Lidar and Samansa Maneshi for helpful
discussions, and we thank Sara Schneider for assistance with the
maximum-likelihood methods. \ We would also like to thank the DARPA-QuIST
program (managed by AFOSR under agreement No. F49620-01-1-0468), the
National Science and Engineering Research Council of Canada, the Canadian
Institute for Photonics Inovation, and eMPOWR for financial support of this
project.

\bibliographystyle{AIP}
\bibliography{process}

Figures:

Figure 1) \ a) A fluorescence image of the state populations in a lattice
obtained by adiabatic decrease of the lattice potential is shown . \ A
stepwise decrease of the potential leads to a clearer separation of the
states as shown in b).

Figure 2) \ \ Matrix of measured projections. \ Input density matrices
(reading left to right: ground state; mixed state; superposition with real
coherence and superposition with imaginary coherence) are shown along the
top while the post-selected states are listed on the side. The table on the
left shows the projections for the input states while the table on the right
shows the corresponding projections after one period. \ Note the decreased
contrast in the $\theta _{x}$ and $\theta _{y}$ projections for the coherent
states. \ All populations are unchanged to within experimental error.

Figure 3) \ The Choi matrix \ after one oscillation in the lattice
characterizing decoherence. \ The left graph shows the real part of the Choi
matrix and the right graph shows the imaginary. \ The matrix is dominated by
real components at the corners. \ The diagonal corners represent the mapping
of populations into populations. \ The off-diagonal corners, which map
coherences into coherences, are significantly less than one, showing a loss
of coherence. \ The dotted lines separates the $2\times 2$ submatrices of
the Choi matrix.

Figure 4) \ Bloch sphere representation of process tomography. \ a) The
initial Bloch sphere representing the space of pure input states. \ b) after
1 period of free evolution the sphere contracts horizontally due to a loss
of coherence. \ c) B.S. after sine drive showing rotation about the y-axis.
\ d) B.S. after cosine drive, showing rotation about the x-axis.

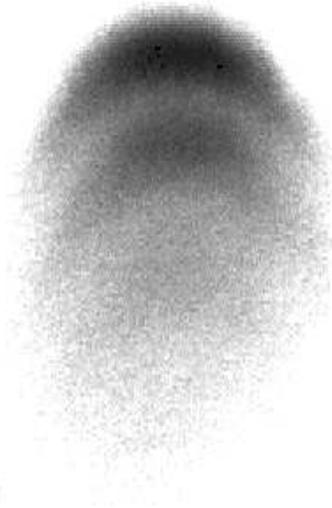
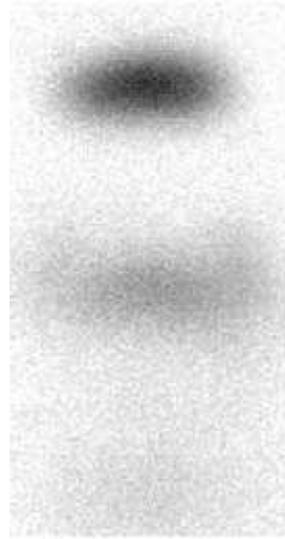
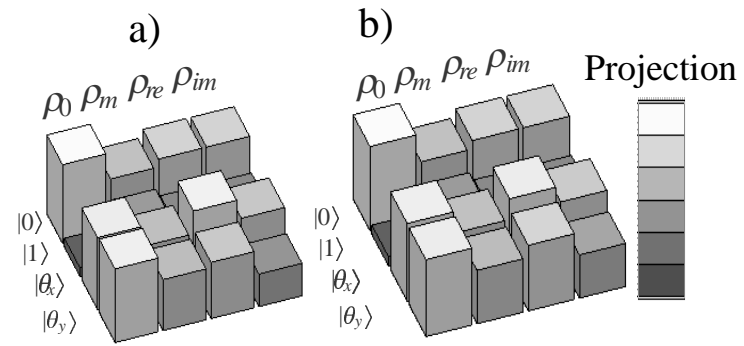
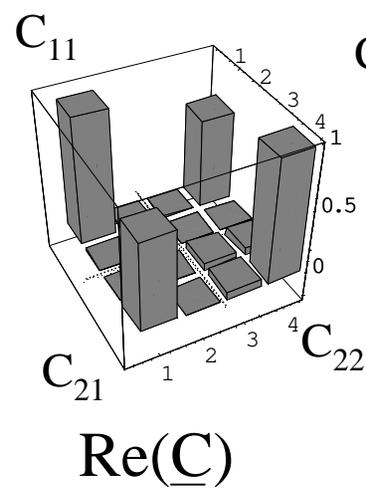
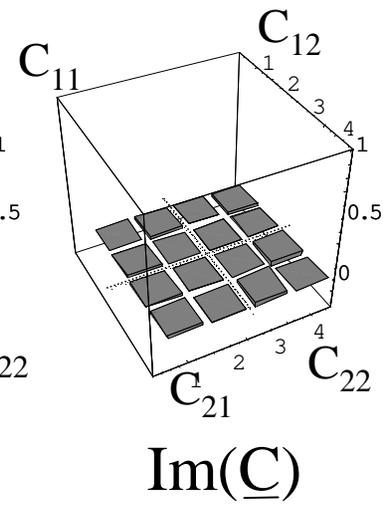
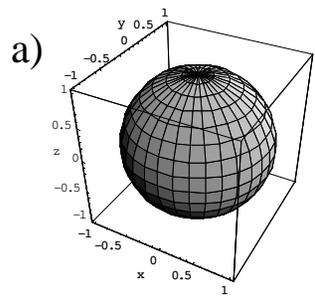
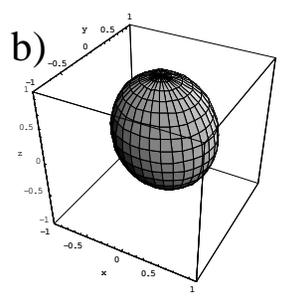
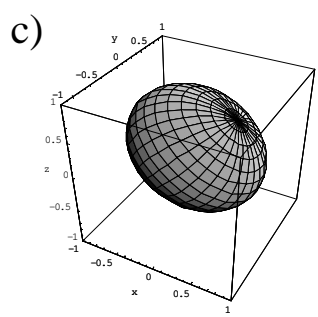
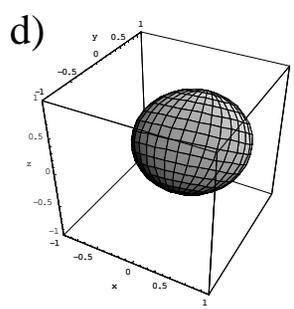
\end{document}